\documentclass[journal]{IEEEtran}
\ifCLASSINFOpdf
\else
\fi

\newtheorem{theorem}{Theorem}[section]

\newtheorem{lemma}[theorem]{Lemma}
\newtheorem{corollary}[theorem]{Corollary}

\newtheorem{exmp}[theorem]{Example}

\newcommand{\EXAMPLE}{\begin{exmp} }
\newcommand{\eoEXAMPLE}{\end{exmp}}

\newcommand{\ignore}[1]{}
\newcommand{\tabincell}[2]{\begin{tabular}{@{}#1@{}}#2 \end{tabular}}
\usepackage{amsmath}
\usepackage{amssymb}
\usepackage{slashbox}
\usepackage[square, comma, sort&compress, numbers]{natbib}

\begin{document}
%
\title{New families of optimal frequency hopping sequence sets}
%
%
%

\author{Jingjun Bao,~Lijun Ji
\thanks{This work was supported by the NSFC under Grants 11222113, 11431003, and a project funded by the priority academic program development of Jiangsu higher education institutions.}
\thanks{ J. Bao and L. Ji are with the Department of Mathematics, Soochow University, Suzhou 215006, P. R. China. E-mail:  baojingjun@hotmail.com; jilijun@suda.edu.cn.}
}

%
%

\markboth{}%
{Shell \MakeLowercase{\textit{et al.}}: Bare Demo of IEEEtran.cls for Journals}
%



\maketitle

\begin{abstract}
Frequency hopping sequences (FHSs) are employed to mitigate the interferences caused by the hits of frequencies in frequency hopping spread spectrum systems.  In this  paper, we present some new algebraic and combinatorial constructions for FHS sets, including an algebraic construction via the linear mapping, two direct constructions by using cyclotomic classes and recursive constructions based on cyclic difference matrices. By these constructions, a number of series of new FHS sets are then produced. These FHS sets are optimal with respect to the Peng-Fan bounds.
\end{abstract}


%
\IEEEpeerreviewmaketitle

\section{Introduction}
%
%
%
%

\IEEEPARstart{F}{requency} hopping (FH) multiple-access is widely used in the modern communication systems such as ultrawideband (UWB), military communications, Bluetooth and so on, for example, \cite{B2003}, \cite{FD1996},
\cite{YG2004}. In FH  multiple-access communication systems, frequency hopping sequences are employed to specify the frequency on which each sender transmits a message at any given time. An important component of FH spread-spectrum systems is a family of sequences having good correlation properties for sequence length over suitable number of available frequencies. The optimality of correlation property is usually measured according to the well-known Lempel-Greenberger bound and Peng-Fan bounds. During these decades, many algebraic or combinatorial constructions for FHSs or FHS sets meeting these bounds have been proposed,
see \cite{CJ2005}-\cite{CY2010}, \cite{DFFJM2009}-\cite{DY2008}, \cite{FMM2004}-\cite{GMY2009}, \cite{RFZ2014}, \cite{YTUP2011}-\cite{ZTPP2011}, and the references therein. Moreover, some generic extension methods have been proposed \cite{CGY2014}, \cite{CHY2009}, \cite{ZCTY2012}, which obtain some new optimal FHS sets.


\newcounter{mytempeqncnt}
\begin{figure*}[!t]
\normalsize
\setcounter{mytempeqncnt}{\value{equation}}

\centerline{\footnotesize KNOWN AND NEW FHS SETS WITH OPTIMA HAMMING CORRELATION}
\smallskip

\begin{center}
\begin{tabular}{|c|c|c|c|c|c|}
\hline
Length & \tabincell{c} {Number of \\ sequences} & $H_{max}$ & \tabincell{c} {Alphabet \\ size} & Constraints & Reference \\
\hline
\tabincell{c}{$p'(p^m-1)$} & $\left\lfloor \frac{a}{p'}\right\rfloor$  & $p'b$ & $a+1$ & \tabincell{c}{$p^m-1=ab$,\\ $a\geq p'(b+1)$, \\gcd$(p',p^m-1)=1$} & \cite{RFZ2014} \\ \hline
\tabincell{c}{$q^m-1$} & $q^u$ & $q^{m-u}$ & $q^u$ & $m > u \geq 1$ & \cite{ZTPP2011} \\ \hline
\tabincell{c}{$\frac{q^m-1}{d}$}  & $d$ & $\frac{q^{m-u}-1}{d}$ & $q^u$ &  \tabincell{c}{ $d|q-1$,\\$ m > u \geq 1$, \\gcd$(d,m)=1$} & \cite{ZTPP2011} \\ \hline
\tabincell{c}{$\frac{q+1}{d}$} & $d(q-1)$ & $1$ & $q$ & $q+1\equiv d \pmod {2d}$ & \cite{DYT2010} \\ \hline
\tabincell{c}{$2^{2t}+1$} & $2^{2t}-1$ & $2^t+1$ & $2^t$ & &\cite{DFFJM2009}\\ \hline
\tabincell{c}{$tp^2$} & $\left\lfloor \frac{p}{t}\right\rfloor$ & $tp$ & $p$& \tabincell{c}{$ p>t\geq 2$}  &  \cite{CHY2009} \\ \hline
\tabincell{c}{$t\frac{q^m-1}{d}$} & $\left\lfloor \frac{d}{t}\right\rfloor$ & $t\frac{q^{m-1}-1}{d}$ & $q$&  \tabincell{c}{$ d\geq t \geq 2,\ d|q-1$,\\ gcd$(m,d)=1$} &  \cite{CHY2009} \\ \hline
\tabincell{c}{$v$} & $f$ & $e$ & $\frac{v-1}{e}+1$ & \tabincell{c}{ $v$ is not a prime or \\ $v$ is a prime with $f\geq e >1$ } & \cite{ZCTY2013} \\ \hline
\tabincell{c}{$p(p^m-1)$}& $p^{u-1}$ & $p^{m-u+1}$ & $p^u$ & $m\geq u>1$  & Theorem\ \ref{optimal p(q^n-1)} \\ \hline
\tabincell{c}{$tv$} & $\lfloor \frac{p_1-1}{t} \rfloor$ & $t$ & $v$ &  $p_1>t>1$  & Theorem\ \ref{optimal tv} \\ \hline
\tabincell{c}{$3v$ } & 2  & 4 & $\frac{3v+1}{4}$ & \tabincell{c}{ each $p_j\equiv 1 \pmod 4$ \\ and $v\not\equiv 0\pmod {25}$ }   & Corollary\ \ref{optimal 3p_1p_2} \\ \hline
\tabincell{c}{$\frac{w(q^m-1)}{d}$}  & $d$& $\frac{q^{m-u}-1}{d}$ & $wq^u$  &  \tabincell{c}{$d|q-1$,\\gcd$(d,m)=1$,\\ $q_1>q^{m-u}$, \\ $m > u \geq 1$ } &  Corollary\ \ref{v(q^n-1)} \\ \hline
\tabincell{c}{$vw$}& $f$ & $e$ & $\frac{v-1}{e}w+\frac{w-1}{e^{'}}+1$ & \tabincell{c}{$e\geq e'\geq 2$ \\ $f\geq 2$, $v\geq e^2$,\\ $q_1\geq p_1$ }  & Theorem\ \ref{optimal v_1v_2} \\ \hline
\tabincell{c}{$vwp'(p^m-1)$}& $\lfloor \frac{a}{p'}\rfloor$
&  $p'b$ & $avw+\frac{(v-1)w}{e}+\frac{w-1}{e^{'}}+1$ & \tabincell{c}{ $p^m-1=ab$,\\ $a\geq p'(b+1)$, \\ $b\geq e\geq e'\geq 2$, \\ gcd$(p',p^m-1)=1$,\\ $q_1 \geq p_1> p^m-1$} & Theorem\ \ref{optimal qv} \\ \hline
\end{tabular}
\end{center}

\hspace{1cm} $q$ is a prime power;

\hspace{1cm} $p,p'$ are primes;

\hspace{1cm} $v$ is an integer with prime factor decomposition $v=p_1^{m_1}p_2^{m_2}\cdots p_s^{m_s}$ with $p_1<p_2<\ldots<p_s$;

\hspace{1cm} $e$ is an integer such that $e|gcd(p_1-1,p_2-1,\ldots,p_s-1)$, and $f=\frac{p_1-1}{e}$;

\hspace{1cm} $w$ is any an integer with prime factor decomposition $w=q_1^{n_1}q_2^{n_2}\cdots q_t^{n_t}$ with $q_1<q_2<\ldots<q_t$;

\hspace{1cm} $e'$ is an integer such that $e'|gcd(q_1-1,q_2-1,\ldots,q_t-1)$;

\hspace{1cm} $t,m, d, u, a$ and $b$ are positive integers.

\setcounter{equation}{\value{mytempeqncnt}}
\vspace*{4pt}
\end{figure*}

In this paper, we present some constructions for FHS sets with optimal Hamming correlations. First of all, we present an algebraic construction for optimal FHS sets by using linear mapping. Secondly, we give two direct constructions for FHS sets by using cyclotomic classes. Finally, we present recursive constructions for FHS sets, which increase their lengths and alphabet sizes, and preserve their maximum Hamming correlations. Our constructions yield optimal FHS sets with new and flexible parameters not covered in the literature. The parameters of FHS sets with optimal Hamming correlations from the known constructions and the new ones are listed in the table.

The remainder of this paper is organized as follows. Section II introduces the known bounds on the Hamming correlations of FHSs and FHS sets. Section III gives an algebraic construction of FHS sets. Section IV  presents two direct constructions and recursive constructions of FHS sets. Section V concludes this paper with some remarks.
\section{Preliminaries}

For any positive integer $l\geq2$, let $F=\{f_0, f_1,\ldots, f_{l-1}\}$ be a set of $l$ available frequencies, also called an {\em alphabet}. A sequence $X=\{x(t)\}_{t=0}^{n-1}$ is called a {\em frequency hopping sequence} (FHS) of
length $n$ over $F$ if $x(t)\in F$ for $0\leq t\leq n-1$. For any two FHSs  $X=\{x(t)\}_{t=0}^{n-1}$ and  $Y=\{y(t)\}_{t=0}^{n-1}$ of length $n$ over $F$, their {\em Hamming correlation} $H_{X,Y}$ is defined by
\begin{equation}
\label{Correlation}
H_{X,Y}(\tau)=\sum_{t=0}^{n-1}h[x(t), y(t+\tau)], 0\leq \tau < n,
\end{equation}
where $h[a,b]=1$ if $a=b$ and $0$ otherwise, and the addition is performed modulo $n$. If $x(t)=y(t)$ for $0\leq t \leq n-1$, i.e., $X=Y$, we call $H_{X,X}(\tau)$ the {\em Hamming autocorrelation} of $X$; otherwise, we say $H_{X,Y}(\tau)$ the {\em Hamming cross-correlation} of $X$ and $Y$. For any two distinct sequences $X$ and $Y$ over $F$, we define
$$H(X)=\max\limits_{1\leq \tau < n}\{H_{X,X}(\tau)\}$$
and
$$H(X,Y)=\max\limits_{0\leq \tau < n}\{H_{X,Y}(\tau)\}.$$

Lempel and Greenberger established the following lower bound on $H(X)$ \cite{LG1974}.

\begin{lemma}\cite{LG1974}\label{FHS BOUND}
For every FHS $X$ of length $n$ over an alphabet of size $l$, it holds that
\begin{equation}
\label{Bound 4}
H(X)\geq \left\lceil\frac{(n-\epsilon)(n+\epsilon-l)}{l(n-1)} \right\rceil,
\end{equation}
where $\epsilon$ is the least nonnegative residue of $n$ modulo $l$.
\end{lemma}

In general, it is more convenient to use a simplified version of the Lempel-Greenberger bound given in Corollary 1.2 of \cite{FMM2004}.

\begin{lemma}\cite{FMM2004}\label{A optimal}
For every FHS $X$ of length $n$ over an alphabet of size $l$, it holds that
\[
\begin{array}{l}
H(X) \geq
\left\{\begin{array}{ll}
k & {\rm  if} \ \ n\neq l,\ \ {\rm and} \\
0  & {\rm  if} \ \ n= l,\
\end{array}
\right .
\end{array}
\]
where $k=\lfloor\frac{n}{l}\rfloor$. This implies that when $n>l$, if $H(X)=k$, then the sequence is optimal.
\end{lemma}

Let $S$ be a set of $M$ FHSs of length $n$ over an alphabet $F$ of size $l$. The maximum nontrivial Hamming correlation $H(S)$ of the sequence set $S$ is defined by
$$H(S)=\max\{\max\limits_{X\in{S}}H(X),\max\limits_{X,Y\in S,\ X\neq Y}H(X,Y)\}.$$

Throughout this paper, we use $(n,M,\lambda;l)$ to denote a set $S$ of $M$ FHSs of length $n$ over an alphabet $F$ of size $l$, where $\lambda=H(S)$. And we use $(n,\lambda;l)$ to denote an FHS $X$ of length $n$ over an alphabet $F$ of size $l$, where $\lambda=H(X)$.

In 2004, Peng and Fan \cite{PF2004} described the following bounds on $H(S)$, which take into consideration the number of sequences in the set $S$.

\begin{lemma}\cite{PF2004}\label{set}
Let $S$ be a set of $M$ sequences of length $n$ over an alphabet $F$ of size $l$.
Define $I=\left\lfloor \frac{nM}{l}\right\rfloor$. Then
\begin{equation}
\label{Bound 5}
H(S)\geq \left\lceil\frac{(nM-l)n}{(nM-1)l}\right\rceil
\end{equation}
and
\begin{equation}
\label{Bound 6}
H(S)\geq  \left\lceil\frac{2InM-(I+1)Il}{(nM-1)M}\right\rceil.
\end{equation}
\end{lemma}

A FHS set is called {\em optimal} if one of the bounds in Lemma \ref{set} is met.
We give a simplified version of the Peng-Fan bound (\ref{Bound 6}) in the following corollary.

\begin{corollary}\label{bounds}
Let $S$ be a set of $M$ sequences of length $n$ over an alphabet $ F$ of size $l$ with $M>1$.
Then
\[
\begin{array}{l}
H(S)\geq
\left\{\begin{array}{ll}
k & {\rm  if} \ \ \epsilon M < l,\ \ {\rm and} \\
k+1  & {\rm otherwise},\
\end{array}
\right .
\end{array}
\]
where $\epsilon$ is the least nonnegative residue of $n$ modulo $l$ and $k=\frac{n-\epsilon}{l}$.
This implies that when $n>l$, if
\[
\begin{array}{l}
H(S)=
\left\{\begin{array}{ll}
k & {\rm  if} \ \ \epsilon M < l,\ \ {\rm and} \\
k+1  & {\rm otherwise},\
\end{array}
\right .
\end{array}
\]
then the FHS set is optimal.
\end{corollary}

\begin{IEEEproof}
Denote $I=\left\lfloor \frac{nM}{l}\right\rfloor$, then $nM=Il+r$,  where $r$ is a nonnegative integer and $r<l$. Simple computation shows that $I=\left\lfloor \frac{nM}{l}\right\rfloor=\left\lfloor \frac{(kl+\epsilon)M}{l}\right\rfloor=kM+\left\lfloor \frac{\epsilon M}{l}\right\rfloor$. Since $-1<\frac{I(r+1-l)}{Il+r-1}\leq0$, we obtain
\[
\begin{array}{l}
\vspace{0.2cm}\left\lceil \frac{2InM-(I+1)Il}{(nM-1)M}\right\rceil=\left\lceil \frac{I(Il+2r-l)}{(Il+r-1)M}\right\rceil
\\ \vspace{0.2cm}\hspace{2.5cm}=\left\lceil \frac{I}{M}+ \frac{I(r+1-l)}{M(Il+r-1)}\right\rceil
\\ \vspace{0.2cm}\hspace{2.5cm}=\left\lceil \frac{I}{M}\right\rceil \\
\vspace{0.2cm}\hspace{2.5cm}=k+\left\lceil \frac{\lfloor \frac{\epsilon M}{l}\rfloor}{M}\right\rceil \\
\vspace{0.2cm}\hspace{2.5cm}=
\left\{\begin{array}{ll}
k & {\rm  if} \ \ \epsilon M < l,\ \ {\rm and} \\
k+1  & {\rm otherwise}.\
\end{array}
\right .
\end{array}
\]
This completes the proof.
\end{IEEEproof}

Our objective is to construct as many FHS sets attaining the bound (4) as possible.
\section{Algebraic Construction} %

In this section, we use a linear mapping from $(GF(p^m),+)$ to $(GF(p^u),+)$ to present an optimal $(p(p^m-1),p^{u-1},p^{m-u+1};p^{u})$-FHS set.

{\bf Construction A}\ \
Let $p$ be a prime and let $u,m$ be two positive integers with $1<u\leq m$. Let $\alpha,\beta$ be primitive elements of $GF(p^m)$
and $GF(p^u)$ respectively. Define a mapping $\sigma$ from $(GF(p^m),+)$ to $(GF(p^u),+)$ by $\sigma(a_0+a_1\alpha+\cdots+a_{m-1}\alpha^{m-1})=a_0+a_1\beta+\cdots+a_{u-1}\beta^{u-1}$ for any $a_0,a_1,\ldots,a_{m-1}\in GF(p)$, and denote $R=\{a_1\alpha+\cdots+a_{u-1}\alpha^{u-1}:a_1,\ldots,a_{u-1}\in GF(p)\}$.
Let $S=\{X^a:~a\in R\}$ be a set of $p^{u-1}$ FHSs of length $p(p^m-1)$ over $GF(p^u)$, where $X^a=\{X^a(t)\}_{t=0}^{p(p^m-1)-1}$ is defined by $X^a(t)=\sigma (\alpha^{\langle t\rangle_{p^m-1}})+ \langle t\rangle_{p}+a$, and $\langle x \rangle_y$ denotes the least nonnegative residue of $x$ modulo $y$ for any positive integer $y$ and any integer $x$.

\begin{theorem}\label{optimal p(q^n-1)} The FHS set $S$ generated by Construction A is an optimal $(p(p^m-1),p^{u-1},p^{m-u+1};p^{u})$-FHS set.
\end{theorem}

\begin{IEEEproof} By the definition of $\sigma$, it is clear that $\sigma$ is linear mapping from $(GF(p^m),+)$ to $(GF(p^u),+)$ of rank $u$.
For any $y\in GF(p^u)$, denote $N_y=\{x\in GF(p^m):\sigma (x)=y\}$. By the linear algebra theory the number of solutions of the equation $\sigma (x)=y$ is $p^{m-u}$. That is $|N_y|=p^{m-u}$ for any $y\in GF(p^u)$.

For $0\leq \tau<p(p^m-1)$ and $a,b\in R$, the Hamming correlation $H_{X^a,X^b}(\tau)$ is given by
\[
\begin{array}{l}
\vspace{0.2cm} H_{X^a,X^b}(\tau)
\\ \vspace{0.2cm}=\sum\limits_{t=0}^{p(p^m-1)-1}h[\sigma (\alpha^{\langle t\rangle_{p^m-1}})+\langle t\rangle_{p}+a, \\
\vspace{0.2cm} \hspace{2.2cm} \sigma (\alpha^{\langle t+ \tau \rangle_{p^m-1}})+  \langle t+\tau \rangle_{p} +b]
\\ \vspace{0.2cm}=\sum\limits_{t=0}^{p(p^m-1)-1}h[a-b-\langle \tau \rangle_{p}, \sigma (\alpha^{\langle t\rangle_{p^m-1}}(\alpha^{\langle  \tau \rangle_{p^m-1}}-1))].
\end{array}
\]
Let $\tau_0=\langle \tau\rangle_{p^m-1}$ and $\tau_1=\langle \tau\rangle_{p}$
According to the values of $a,b$ and $\tau$, we compute $H_{X^a,X^b}(\tau)$ in four cases.

Case 1: $a=b$ and $\tau_1=0$. In this case $\tau_0\neq 0$ and $\alpha^{\tau_0}-1\in GF(p^m)\setminus \{0\}$. For $x\in GF(p^m)\setminus \{0\}$, let $t_x$ be an integer such that $\alpha^{\langle t_x\rangle_{p^m-1}}(\alpha^{\tau_0}-1)=x$. Then \[
\begin{array}{l}
\vspace{0.2cm}H_{X^a,X^a}(\tau)\\  \vspace{0.2cm} =\sum\limits_{t=0}^{p(p^m-1)-1}h[0, \sigma (\alpha^{\langle t\rangle_{p^m-1}}(\alpha^{\tau_0}-1))]
\\ \vspace{0.2cm}=\sum\limits_{x\in N_{0}\setminus\{0\}}|\{t\equiv t_x \pmod{p^m-1}:~ 0\leq t<p(p^m-1) \}|
\\ \vspace{0.2cm}=(p^{m-u}-1)p.
\end{array}
\]

Case 2: $a=b$ and $\tau_1\neq 0$. If $\tau_0=0$, then
\[
\begin{array}{l}
H_{X^a,X^a}(\tau)=\sum\limits_{t=0}^{p(p^m-1)-1}h[-\tau_1, 0]=0.
\end{array}
\]
Otherwise, let $t_x$ be an integer such that $\alpha^{\langle t_x\rangle_{p^m-1}}(\alpha^{\tau_0}-1)=x$ for any $x\in GF(p^m)\setminus \{0\}$. Then
\[
\begin{array}{l}
\vspace{0.1cm}H_{X^a,X^a}(\tau)
\\ \vspace{0.1cm} =\sum\limits_{t=0}^{p(p^m-1)-1}h[-\tau_1, \sigma (\alpha^{\langle t\rangle_{p^m-1}}(\alpha^{\tau_0}-1))]
\\ \vspace{0.1cm}=\sum\limits_{x\in N_{-\tau_1}}|\{t\equiv t_x \pmod{p^m-1}: 0\leq t< p(p^m-1)\}|
\\ \vspace{0.1cm}=p^{m-u+1}.
\end{array}
\]

Case 3: $a\neq b$ and $\tau_0=0$. Since $a-b\notin GF(p)$ , we have $a-b\neq \tau_1$. Then $$H_{X^a,X^b}(\tau)=\sum\limits_{t=0}^{p(p^m-1)-1}h[a-b-\tau_1, 0]=0.$$

Case 4: $a\neq b$ and $\tau_0\neq 0$. For $x\in GF(p^m)\setminus \{0\}$, let $t_x$ be an integer such that $\alpha^{\langle t_x\rangle_{p^m-1}}(\alpha^{\tau_0}-1)=x$. Since $a-b\neq \tau_1$, we have
\[
\begin{array}{l}
\vspace{0.1cm} H_{X^a,X^b}(\tau) \vspace{0.2cm}
\\ \vspace{0.1cm} =\sum\limits_{t=0}^{p(p^m-1)-1}h[a-b-\tau_1, \sigma (\alpha^{\langle t\rangle_{p^m-1}}(\alpha^{\tau_0}-1))]
\\ \vspace{0.1cm}=\sum\limits_{x\in N_{a-b-\tau_1}}|\{t\equiv t_x \pmod{p^m-1}:~0\leq t< p(p^m-1)\}|
\\ \vspace{0.1cm}=p^{m-u+1}.
\end{array}
\]
Thus, $S$ is a $(p(p^m-1),p^{u-1},p^{m-u+1};p^u)$-FHS set. It remains to prove that this FHS set is also optimal.

Since $k=\left\lfloor \frac{n}{l}\right\rfloor=\left\lfloor \frac{p(p^m-1)}{p^u}\right\rfloor=p^{m+1-u}-1$, we have that
\[
\begin{array}{l}
\vspace{0.1cm}\epsilon=n-kl \\
\vspace{0.1cm} \hspace{0.2cm}=p(p^m-1)-(p^{m-u+1}-1)p^{u}\\
\vspace{0.1cm} \hspace{0.2cm}=p^{u}-p, \ {\rm and}\\
\vspace{0.1cm}\epsilon M-l
\\ \vspace{0.1cm} =p^{u-1}(p^{u}-p)-p^{u}\\
\vspace{0.1cm} =(p^{u-1}-2)p^{u}\geq 0.
\end{array}
\]
By Corollary \ref{bounds}, $S$ is also optimal. This completes the proof.
\end{IEEEproof}

\section{Combinatorial Constructions} %

\subsection{Combinatorial Characterization of FHS Sets} %


Following \cite{GMY2009}, we describe a connection between FHS sets and partition-type BNCDPs in this subsection.

Throughout this paper we always assume that $\mathbf{I}_l=\{0,1,2,\ldots,l-1\}$ and $\mathbb{Z}_n$ is the residual-class ring of integers modulo $n$.

An $(n,\lambda;l)$-FHS, $X=(x(0),x(1),\ldots,x(n-1))$, over a frequency library $F$, can be interpreted as a family of $l$ sets $B_0,B_1,\ldots,B_{l-1}$ such that each set $B_i$ corresponds to frequency $i\in F$ and the elements in each set $B_i$ specify the position indices in the FHS $X$ at which frequency $i$ appears.

Associated with a non-empty subset $B \subseteq \mathbb{Z}_n$, the difference list of $B$ from combinatorial design theory is defined to be the multiset
  $$ \Delta (B) = \left\{a-b:~ a, b \in B~ \mbox {and}~  a \ne b\right\}.$$
For any family ${\cal B} = \{B_{0}, B_{1}, \ldots, B_{l-1}\}$
 of $l$ non-empty subsets (called {\em base blocks}) of $\mathbb{Z}_n$, define the difference list of ${\cal B}$ to be the union of multisets

 $$ \Delta ({\cal B}) = \displaystyle{\bigcup_{i\in \mathbf{I}_l}\Delta (B_i)}.$$
If the difference list $\Delta({\cal B})$ contains each non-zero residue of $\mathbb{Z}_n$ at most $\lambda$ times, then ${\cal B}$ is said to be an $(n, K, \lambda)$-CDP ({\em cyclic difference packing}), where $K=\{|B_i|: i\in \mathbf{I}_l\}$. The number $l$ of the base blocks in ${\cal B}$ is refereed to as the {\em size} of the CDP. If the difference list $\Delta({\cal B})$ contains each non-zero residue of $\mathbb{Z}_n$ exactly $\lambda$ times, then ${\cal B}$ is said to be an $(n, K, \lambda)$-CDF ({\em  cyclic difference family}). When $K=\{k\}$, we simply write $k$ for $\{k\}$.

An $(n,K,\lambda)$-CDP with a collection of blocks ${\cal B}=\{B_0,B_1,\ldots,B_{l-1}\}$ is called a {\em partition-type cyclic difference packing} if every element of $\mathbb{Z}_n$ is contained in exactly one base block of ${\cal B}$.

In 2004, Fuji-Hara et al. \cite{FMM2004} revealed a connection between FHSs and partition-type cyclic difference packings as follows.
\begin{theorem}\cite{FMM2004}\label{FHS=DP}
There exists an $(n,\lambda;l)$-FHS over a frequency library $F$ if and only if there exists a partition-type $(n,K,\lambda)$-CDP of size $l$, ${\cal B}=\{B_0,B_1,\ldots,B_{l-1}\}$ over $\mathbb{Z}_n$, where $K=\{|B_i|:~0\leq i\leq l-1\}$.
\end{theorem}

Let $A,B$ be two non-empty subsets of $\mathbb{Z}_n$.  {\em The list of external difference} of ordered pair $(A,B)$ is the multiset $$\Delta_E(A,B)=\{y-x:~(x,y)\in A\times B\}.$$  Note that the list of external difference $\Delta_E(A,B)$ may contain zero. For any residue $d\in \mathbb{Z}_n$, the number of occurrences of $d$ in $\Delta_E(A,B)$ is clearly equal to $|(A+d)\cap B|$.

Let ${\cal B}_j, 0\leq j\leq M-1$, be a collection of $l$ subsets of $\mathbb{Z}_n$ where ${\cal B}_j=\{B_0^j, \ldots,B_{l-1}^j \}$. The list of external difference of ordered pair $({\cal B}_j,{\cal B}_{j'})$, $0\leq j\neq j'<M$, is the union of multisets
$$\Delta_E({\cal B}_j,{\cal B}_{j'})= \bigcup\limits_{i\in \mathbf{I}_l} \Delta_E(B_i^j,B_i^{j'}).$$
The set $\{{\cal B}_0, \ldots, {\cal B}_{M-1}\}$ of CDPs is said to be
{\em balanced nested} with index $\lambda$ and denoted by $(n,\{K_0,\ldots, K_{M-1}\},\lambda)$-BNCDP
if each ${\cal B}_j$ is an $(n,K_j,\lambda)$-CDP of size $l$, and $\Delta_E({\cal B}_j, {\cal B}_{j'})$
contains each residue of $\mathbb{Z}_n$ at most $\lambda$ times for any $j\neq j'$. If each ${\cal B}_j$ is a partition-type CDP, then the $(n,\{K_0,\ldots, K_{M-1}\},\lambda)$-BNCDP is called  {\em partition-type}. For convenient, the number $l$ of the base blocks in ${\cal B}_j$ is also said to be the size of the BNCDP.

In 2009, Ge et al. \cite{GMY2009} revealed a connection between FHS sets and partition-type BNCDPs as follows.

\begin{theorem}\cite{GMY2009} \label{FHS sets}
There exists an $(n,M,\lambda;l)$-FHS set over a frequency library $F$ if and only if there exists a partition-type $(n,\{K_0,\ldots,K_{M-1}\},\lambda)$-BNCDP of size $l$, $\{{\cal B}_j:0\leq j< M\}$ over $\mathbb{Z}_n$, where ${\cal B}_j=\{B_0^j,\ldots, B_{l-1}^j\}$ and $K_j=\{|B^j_i|:0\leq i<l\}$  for $0\leq j < M$.
\end{theorem}

\subsection{ Direct Constructions of Optimal FHS Sets} %
In this subsection, we give two direct constructions for FHS sets.

Let $A$ be a subset of $\mathbb{Z}_v$, we define $\lambda\cdot[A]=\bigcup\limits_{i=0}^{\lambda-1}A$, where $\bigcup$ denotes the multiset union. Define by $U(\mathbb{Z}_v)$ the set of all units in $\mathbb{Z}_v$.

\begin{theorem}\label{optimal tv}
Let $v$ be an odd integer and $p_1$ the least prime divisor of $v$. For any integer $t$ with $1<t<p_1$, there exists an optimal $(tv,\left\lfloor \frac{p_1-1}{t}\right\rfloor,t;v)$-FHS set.
\end{theorem}

\begin{IEEEproof}
Denote $a=\left\lfloor \frac{p_1-1}{t}\right\rfloor$ and write $v$ as $p_1^{m_1}p_2^{m_2}\cdots p_s^{m_s}$ for $s$ positive integers $m_1,m_2,\ldots,m_s$ and $s$ distinct primes $p_1,p_2,\ldots,p_s$.
 For $0\leq i\leq ta-1$, by the Chinese Remainder Theorem, there exists a unique element $\theta_i$ such that $$\theta_i\equiv i+1\pmod {p_j^{m_j}}\ {\rm for}\ 1 \leq j \leq s.$$  Clearly, for $0\leq x\neq y< ta$, $\theta_x\ {\rm and}\ \theta_x-\theta_y$ belong to $U(\mathbb{Z}_v)$.

By Theorem \ref{FHS sets}, we need to construct a partition-type $(tv,\{K_0,\ldots,K_{a-1}\},t)$-BNCDP of size $v$ over $\mathbb{Z}_{tv}$.

For $0\leq u<a$ and $0\leq c<v$, set
\[
\begin{array}{l}
\vspace{0.2cm}\ B_{c}^u=\{b+ct\theta_{b+ut} \pmod {tv}:~ 0\leq b<t \}, \ {\rm and}\ \\
\vspace{0.2cm}\ {\cal B}_u=\{B_{c}^u :~ 0\leq c<v \}.
\end{array}
\]

Firstly, we show that each ${\cal B}_u$ is a partition-type $(tv,t,t)$-CDP.

Since $\theta_{b+ut}\in U(\mathbb{Z}_{v})$ for $0\leq b<t$, we have that
\[
\begin{array}{l}
\vspace{0.2cm} \bigcup\limits_{0\leq c<v} B_{c}^u=\bigcup\limits_{0\leq c<v} \bigcup\limits_{0\leq b<t} \{b+ct\theta_{b+ut} \}\\
\vspace{0.2cm} \hspace{1.5cm} =  \bigcup\limits_{0\leq b<t} \{b+ct :~ 0\leq c<v \} \\
\vspace{0.2cm} \hspace{1.5cm} = \mathbb{Z}_{tv}. \\
\end{array}
\]
Thus, ${\cal B}_u$ is a partition of $\mathbb{Z}_{tv}$.

Since $\theta_{j+ut}-\theta_{b+ut}\in U(\mathbb{Z}_v)$ for $0\leq b<j<t$,
we get
\[
\begin{array}{l}
\vspace{0.2cm} \Delta({\cal B}_u)=\bigcup\limits_{c=0}^{v-1}\Delta(B_{c}^u)
\\ \vspace{0.2cm}=\bigcup\limits_{c=0}^{v-1}\{\pm (j-b+ ct(\theta_{j+ut}-\theta_{b+ut})):~ 0\leq b<j<t\}
\\ \vspace{0.2cm}=\{ \pm(j-b)+ct:~ 0\leq b<j<t,0\leq c<v\}
\\ \vspace{0.2cm}=t[\mathbb{Z}_{tv}\setminus t\mathbb{Z}_{tv}],
\end{array}
\]
where $t\mathbb{Z}_{tv}=\{0,t,2t,\ldots,(v-1)t\}$.
Hence, ${\cal B}_u$ is a partition-type $(tv,t,t)$-CDP for $0\leq u< a$.

Secondly, we show that $\Delta_E( {\cal B}_u,{\cal B}_{u'})$ contains each element of $ \mathbb{Z}_{tv}$ at most $t$ times for $0\leq u\neq u'< a$.

Since $\theta_{j+u't}-\theta_{b+ut}\in U(\mathbb{Z}_v)$ for $0\leq b,j<t$,
we get
\[
\begin{array}{l}
\vspace{0.2cm} \Delta_E({\cal B}_u, {\cal B}_{u'})=\bigcup\limits_{c=0}^{v-1}\Delta_E(B_{c}^u, B_{c}^{u'})
\\ \vspace{0.2cm}=\bigcup\limits_{c=0}^{v-1}\{j-b + ct(\theta_{j+u't}-\theta_{b+ut}):~ 0\leq b,j<t\}
\\ \vspace{0.2cm}=\{j-b+ct:~ 0\leq b,j<t,\ 0\leq c<v\}
\\ \vspace{0.2cm}=t[\mathbb{Z}_{tv}].
\end{array}
\]
It follows that $\{{\cal B}_u:~0\leq u <a\}$ is a partition-type $(tv,\{K_0,\ldots,K_{a-1}\},t)$-BNCDP of size $v$.

Finally, applying Theorem \ref{FHS sets} we obtain a $(tv,\left\lfloor \frac{p_1-1}{t}\right\rfloor,t;v)$-FHS set.
Since $\epsilon=0$, by Corollary \ref{bounds} this FHS set is also optimal.
\end{IEEEproof}

\begin{theorem}{\label{optimal 3p}}
For any prime $p\equiv1 \pmod4$, there exists an optimal $(3p,2,4;\frac{3p+1}{4})$-FHS set.
\end{theorem}
\begin{IEEEproof}
Let $\alpha$ be a primitive element in $\mathbb{Z}_p$ and let $t=\frac{p-1}{4}$. Then
\begin{equation}\label {3}
\frac{\alpha^{t}-1}{\alpha^{t}+1}=\frac{\alpha^{t}+\alpha^{2t}}{\alpha^{t}+1}=\alpha^t
\end{equation}
and
\begin{equation}\label {4}
\bigcup\limits_{0\leq i< t}\{\alpha^i,-\alpha^i,\alpha^{i+t},-\alpha^{i+t}\}=\mathbb{Z}_p\setminus\{0\}.
\end{equation}

Since gcd$(3,p)=1$, we have that $\mathbb{Z}_{3p}$ is isomorphic to $\mathbb{Z}_3\times \mathbb{Z}_p$.
Let
\[
\begin{array}{l}
\vspace{0.2cm} A_0^i=\{(0,\alpha^i),(0,-\alpha^i),(1,\alpha^{i+t}),(1,-\alpha^{i+t})\},\\
\vspace{0.2cm} A_1^i=A_0^i+(1,0), A_2^i=A_0^i+(2,0),\vspace{0.1cm}\\
\vspace{0.2cm} B_0^i=\{(0,\alpha^{i+1}),(0,-\alpha^{i+1}),(1,\alpha^{i+t+1}),(1,-\alpha^{i+t+1})\},\\
\vspace{0.2cm} B_1^i=B_0^i+(2,0), B_2^i=B_0^i+(1,0),
\end{array}
\]
where $0\leq i< t$. Set
\[
\begin{array}{l}
\vspace{0.2cm} {\cal B}_0=\{A_j^i: 0\leq i< t, 0\leq j<3\}\cup \{\mathbb{Z}_3 \times \{0\}\},\\
{\cal B}_1=\{B_j^i: 0\leq i< t, 0\leq j<3\}\cup \{ \mathbb{Z}_3 \times \{0\}\}.\\
\end{array}
\]
In view of equality (\ref{4}), each ${\cal B}_i$ ($0\leq i<2$) is a partition of $\mathbb{Z}_3\times\mathbb{Z}_p$.
Next we show that $\{ {\cal B}_0,{\cal B}_1 \}$ is a $(3p,\{\{3,4\},\{3,4\}\},4)$-BNCDP.

Firstly, we show that each ${\cal B}_i$ is a $(3p,\{3,4\},3)$-CDP of size $\frac{3p+1}{4}$.

It is straightforward that
\[
\begin{array}{l}
 \vspace{0.2cm} \Delta({\cal B}_0)=\bigcup \limits_{z=0}^2\{z\} \times \Delta_z, \ {\rm where} \\
 \vspace{0.2cm} \Delta_0= 3[ \bigcup \limits_{i=0}^{t-1} \{ \pm 2\alpha^i,\pm 2\alpha^{i+t} \}], \\
 \vspace{0.2cm} \Delta_1= 3 [\bigcup\limits_{i=0}^{t-1} \{\pm \alpha^{i}(\alpha^{t}-1), \pm \alpha^{i}(\alpha^{t}+1)\}\cup\{0\}] , \\
 \Delta_2= -\Delta_1.
 \end{array}
\]
In view of equality (\ref{3}) and equality (\ref{4}), we have
\[
\begin{array}{l}
 \vspace{0.2cm} \Delta({\cal B}_0)
=3[ \mathbb{Z}_3\times\mathbb{Z}_p\setminus \{(0,0)\}].
\end{array}
\]
Similarly, we have $\Delta({\cal B}_1)=3[ \mathbb{Z}_3\times\mathbb{Z}_p\setminus \{(0,0)\}]$. Thus, each ${\cal B}_i$ is a $(3p,\{3,4\},3)$-CDP.

Secondly, we show that $\Delta_E( {\cal B}_0,{\cal B}_1)$ contains each element of $ \mathbb{Z}_3 \times \mathbb{Z}_p$ at most four times.

It is straightforward that
\[
\begin{array}{l}
 \vspace{0.2cm} \Delta_E({\cal B}_0,{\cal B}_1)\\
\vspace{0.2cm} =\left(\bigcup \limits_{i=0}^{t-1}\bigcup \limits_{j=0}^2\Delta_E(A^i_j,B^i_j)\right)\bigcup \Delta_E(\mathbb{Z}_3 \times \{0\}, \mathbb{Z}_3 \times \{0\})\\
\vspace{0.2cm} =\bigcup \limits_{z=0}^2\{z\} \times \Delta_z, \ {\rm where} \\
\vspace{0.2cm}  \Delta_0= \bigcup \limits_{0\leq i< t} \{ \pm \alpha^i(1-\alpha),\pm \alpha^i(1+\alpha),\pm \alpha^{i+t}(1-\alpha),\\
\vspace{0.25cm} \hspace{1.9cm} \pm \alpha^{i+t}(\alpha+1) ,\pm(\alpha^{t}-\alpha)\alpha^{i},\pm(\alpha^{t}+\alpha)\alpha^{i}, \\
\vspace{0.25cm} \hspace{1.9cm}  \pm(1-\alpha^{t+1})\alpha^{i}, \pm (1+\alpha^{t+1})\alpha^{i} \}\cup\{0,0,0\}, \\
 \Delta_2= \Delta_1=\Delta_0.
 \end{array}
\]
In view of equality (\ref{4}), we have that
\[
\begin{array}{l}
 \bigcup \limits_{0\leq i< t} \{ \pm \alpha^i(1-\alpha),\pm \alpha^{i+t}(1-\alpha)\}=\mathbb{Z}_p\setminus \{0\}\ {\rm and}\\
 \bigcup \limits_{0\leq i< t} \{ \pm \alpha^i(1+\alpha),\pm \alpha^{i+t}(\alpha+1)\}=\mathbb{Z}_p\setminus \{0\}.
\end{array}
\]
Since $\frac{\alpha^{t}-\alpha}{1+\alpha^{t+1}}=\frac{\alpha^{t}+\alpha^{2t+1}}{1+\alpha^{t+1}}=\alpha^{t}$ and  $\frac{\alpha^{t}+\alpha}{1-\alpha^{t+1}}=\frac{\alpha^{t}-\alpha^{2t+1}}{1-\alpha^{t+1}}=\alpha^{t}$, we get
\[
\begin{array}{l}
 \bigcup \limits_{0\leq i< t} \{ \pm(\alpha^{t}-\alpha)\alpha^{i},\pm (1+\alpha^{t+1})\alpha^{i}\}=\mathbb{Z}_p\setminus \{0\},\ {\rm and}\\
 \bigcup \limits_{0\leq i< t} \{ \pm(\alpha^{t}+\alpha)\alpha^{i}, \pm(1-\alpha^{t+1})\alpha^{i}\}=\mathbb{Z}_p\setminus \{0\}.
\end{array}
\]
Thus, $\Delta_E({\cal B}_0,{\cal B}_1)=4[\mathbb{Z}_3\times(\mathbb{Z}_p\setminus \{0\})]\cup 3[\{(0,0),(1,0),(2,0)\}]$. It follows that $\{{\cal B}_0,{\cal B}_1\}$ is our required BNCDP. By Theorem \ref{FHS sets}, there exists a $(3p,2,4;\frac{3p+1}{4})$-FHS set.

Finally, we show that such an FHS set is optimal.

Since $k=\left\lfloor\frac{n}{l}\right\rfloor=\left\lfloor\frac{3p}{\frac{3p+1}{4}}\right\rfloor=3$, we get $\epsilon = 3p-3\times \frac{3p+1}{4}=\frac{3p-3}{4}$ and $\epsilon M-l=2\times \frac{3p-3}{4}-\frac{3p+1}{4}=\frac{3p-7}{4}>0.$ By Corollary \ref{bounds}, this FHS set is optimal. This completes the proof.
\end{IEEEproof}

\subsection{Recursive constructions of optimal FHS sets}

In this subsection, recursive constructions are used to construct optimal FHS sets. These recursive constructions are based on cyclic difference matrices (CDMs).

A $(w, t, 1)$-CDM is a $t \times  w$ matrix $D=(d_{ij})$ ($0 \le i \le t-1$, $0 \le j \le w-1)$ with entries from $\mathbb{Z}_w$ such that, for any two distinct rows $R_{r}$ and $R_h$, the vector difference $R_h-R_{r}$ contains every residue of $\mathbb{Z}_w$ exactly once. It is easy to see that the property
of a difference matrix is preserved even if we add any element of $\mathbb{Z}_w$ to all entries in any row or column of the difference matrix.
Then, without loss of generality, we can assume that all entries in the first row are zero. Such a difference matrix is said to be {\em normalized}. The $(w,t-1,1)$-CDM obtained from a normalized $(w,t,1)$-CDM by deleting the first row is said to be {\em homogeneous}. The existence of a homogeneous $(w,t-1,1)$-CDM is equivalent to that of a $(w,t,1)$-CDM. Observe that cyclic difference matrices have been extensively studied. A large number of known $(w, t, 1)$-CDMs are well documented in \cite{CD2007}. In particular, the multiplication table of the prime field $\mathbb{Z}_p$ is a $(p, p, 1)$-CDM. By using the usual product construction of CDMs, we have the following existence result.

\begin{lemma}
\label{CDM}\cite{CD2007}
Let $w$ and $t$ be integers with $w\geq t\geq 3$. If $w$ is odd and the least prime factor of $w$ is not less than $t$, then there exists a $(w,t,1)$-CDM.
\end{lemma}

In 2009, Ge et al. \cite{GMY2009} used cyclic difference matrices to established a recursive construction for partition-type BNCDPs  so as to give a recursive construction of FHS sets.
We generalize this construction via balanced nested cyclic relative difference packings.

An $(mg, g, K, \lambda)$-{\em cyclic relative difference packing} (briefly CRDP) is an $(mg,K,\lambda)$-CDP ${\cal B}$ over $\mathbb{Z}_{mg}$ such that $\Delta({\cal B})$ contains each element of $\mathbb{Z}_{mg}\setminus m\mathbb{Z}_{mg}$ at most $\lambda$ times and no element of $m\mathbb{Z}_{mg}$ occurs, where $m\mathbb{Z}_{mg}=\{0,m,\ldots,mg-m\}$.

Let ${\cal B}_j=\{B_0^j,B_1^j,\ldots,B_{u-1}^j\}$ be an $(mg,g,K_j,\lambda)$-CRDP over $\mathbb{Z}_{mg}$ for $0\leq j\leq M-1$. The set $\{{\cal B}_0,\ldots,{\cal B}_{M-1}\}$ is referred as an $(mg,g,\{K_0,K_1,\ldots,K_{M-1}\},\lambda)$-BNCRDP ({\em balanced nested cyclic relative difference packing}) over $\mathbb{Z}_{mg}$ if $\Delta_E({\cal B}_j,{\cal B}_{j'})$ contains each element of $\mathbb{Z}_{mg}\setminus m\mathbb{Z}_{mg}$ at most $\lambda$ times and no element of $m\mathbb{Z}_{mg}$ occurs for any $j\neq j'$. 
For convenience, the size $u$ of ${\cal B}_j$ is also said to be the size of the BNCRDP.

One importance of BNCRDP is that we can put an appropriate BNCDP on its subgroup to derive a new BNCDP.

\begin{lemma}\label{BNCRDP=>BNCDP}
Suppose there exists an $(mg,g,\{K_0,\ldots,K_{M-1}\},\lambda)$-BNCRDP of size $u$ such that each $(mg,g,K_j,\lambda)$-CRDP is a partition of $\mathbb{Z}_{mg}\setminus m\mathbb{Z}_{mg}$ for $0\leq j < M$. If there exists a partition-type $(g,\{K_0',\ldots,K_{M-1}'\},\lambda)$-BNCDP of size $r$ over $\mathbb{Z}_{g}$, then there exists a partition-type $(mg,\{K_0\cup K_0',\ldots,K_{M-1}\cup K_{M-1}'\},\lambda)$-BNCDP of size $u+r$ over $\mathbb{Z}_{mg}$.
\end{lemma}

\begin{IEEEproof}
Let $\{{\cal B}_0, \ldots, {\cal B}_{M-1}\}$ be a given $(mg,g,\{K_0,\ldots,K_{M-1}\},\lambda)$-BNCRDP over $\mathbb{Z}_{mg}$, where $ {\cal B}_j=\{B_i^j:0\leq i<u\}$ for $0\leq j < M$. Let $\{{\cal A}_0, \ldots, {\cal A}_{M-1}\}$ be a partition-type $(g,\{K_0',\ldots,K_{M-1}'\},\lambda)$-BNCDP over $\mathbb{Z}_{g}$, where $ {\cal A}_j=\{A_i^j:0\leq i<r\}$ for $0\leq j <M$. For $0\leq j< M$, set $P_j=\{mA_i^j:~0\leq i<r\}$ and $T_j={\cal B}_j\cup P_j$. Since ${\cal A}_j$ is a partition of $\mathbb{Z}_{g}$, we have that $P_j$ is a partition of $m\mathbb{Z}_{mg}$, consequently, $T_j$ is a partition of $\mathbb{Z}_{mg}$. It remains to prove that $\{T_j:0\leq j< M\}$ is an $(mg,\{K_0\cup K_0',\ldots,K_{M-1}\cup K_{M-1}'\},\lambda)$-BNCDP over $\mathbb{Z}_{mg}$.

On one hand, since $\{{\cal B}_j:~0\leq j < M\}$ is an $(mg,g,\{K_0,\ldots,K_{M-1}\},\lambda)$-BNCRDP over $\mathbb{Z}_{mg}$, we have that $\Delta({\cal B}_j)$ contains each residue of $\mathbb{Z}_{mg}\setminus m\mathbb{Z}_{mg}$ at most $\lambda$ times and no element of $m\mathbb{Z}_{mg}$ occurs.  Since $\{{\cal A}_j:~0\leq j< M\}$ is a $(g,\{K'_0,\ldots,K'_{M-1}\},\lambda)$-BNCDP over $\mathbb{Z}_{g}$, we have that $\Delta(P_j)=m\Delta({\cal A}_j)$ contains each non-zero element of $ m\mathbb{Z}_{mg}$ at most $\lambda$ times. Thus, $\Delta(T_j)$ contains each non-zero element of $\mathbb{Z}_{mg}$ at most $\lambda$ times, $T_j$ is a partition-type $(mg,K_{j}\cup K_{j}', \lambda)$-CDP of size $u+r$. On the other hand, by the definition of BNCRDP, we have that $\Delta_E({\cal B}_j,{\cal B}_{j'})$ contains each residue of $\mathbb{Z}_{mg}\setminus m\mathbb{Z}_{mg}$ at most $\lambda$ times and no element of $m\mathbb{Z}_{mg}$ occurs, and $\Delta_E(P_j,P_{j'})=m\Delta_E({\cal A}_j,{\cal A}_{j'})$ contains each element of $ m\mathbb{Z}_{mg}$ at most $\lambda$ times for $0\leq j\neq j'< M$. Hence, $\Delta_E(T_j,T_{j'})=\Delta_E({\cal B}_j,{\cal B}_{j'})\cup m\Delta_E({\cal A}_j,{\cal A}_{j'})$ contains each element of $\mathbb{Z}_{mg}$ at most $\lambda$ times for $0\leq j\neq j'< M$. It follows that $\{T_j:~0\leq j< M\}$ is an $(mg,\{K_0\cup K_0',\ldots,K_{M-1}\cup K_{M-1}'\},\lambda)$-BNCDP over $\mathbb{Z}_{mg}$. This completes the proof.
\end{IEEEproof}

\begin{theorem}\label{HFHS}
Assume that $\{{\cal B}_0,\ldots,{\cal B}_{M-1}\}$ is an $(mg,g,\{K_0,\ldots, K_{M-1}\},\lambda)$-BNCRDP of size $u$ such that each ${\cal B}_{j}$ is a partition of $\mathbb{Z}_{mg}\setminus m\mathbb{Z}_{mg}$, where ${\cal B}_{j}=\{B^j_i:~0\leq i<u\}$ for $0\leq j < M$. If there exists a homogeneous $(w,t,1)$-CDM over $\mathbb{Z}_w$ with $t=\max \limits_{0\leq i <u}\{\sum \limits_{j=0}^{M-1} |B_i^j|\}$, then there exists an $(mgw,gw,\{K_0,\ldots, K_{M-1}\},\lambda)$-BNCRDP of size $uw$, $\{{\cal B}'_0,\ldots,{\cal B}'_{M-1}\}$, such that each ${\cal B}'_{j}$ is a partition of $\mathbb{Z}_{mgw}\setminus m\mathbb{Z}_{mgw}$ for $0\leq j < M$.

\end{theorem}

\begin{IEEEproof}
Let $\Gamma=(\gamma_{i,j})$ be a given homogeneous $(w,t,1)$-CDM over $\mathbb{Z}_w$. For each collection of the following $M$ blocks
\[
\begin{array}{l}
\vspace{0.2cm} B_i^0=\{a_{i,0,1},\ldots,a_{i,0,k_0}\},
\\ \vspace{0.2cm} B_i^1=\{a_{i,1,k_0+1},\ldots,a_{i,1,k_1}\},
\\ \vspace{0.2cm} \hspace{2cm} \vdots
\\ \vspace{0.2cm} B_i^{M-1}=\{a_{i,M-1,k_{M-2}+1},\ldots,a_{i,M-1,k_{M-1}}\},
\end{array}
\]
where $0\leq i <u$, we construct the following $uw$ new blocks:
\[
\begin{array}{l}
 \vspace{0.2cm}B_{(i,s)}^j=\{a_{i,j,k_{j-1}+1}+mg\gamma_{k_{j-1}+1,s},\ldots,a_{i,j,k_{j}}^j+mg\gamma_{k_{j},s}\}, \\
 \hspace{1.5cm} {\rm where}\ 0\leq j< M, 0\leq s < w.
\end{array}
\]
Set 
\[
\begin{array}{l}
 \vspace{0.2cm} {\cal B}'_j=\{B_{(i,s)}^j:~0\leq i<u, 0\leq s<w\},\\
 \vspace{0.2cm} {\cal B}'=\{{\cal B}'_j:~0\leq j < M \}.\\
\end{array}
\]  
It is left to show that ${\cal B}'$ is the required BNCRDP. 
 
  Since ${\cal B}_j$ is a partition of $\mathbb{Z}_{mg}\setminus m\mathbb{Z}_{mg}$ and each row of $\Gamma$ is a permutation of $\mathbb{Z}_w$, we obtain that ${\cal B}'_j$ is a partition of $\mathbb{Z}_{mgw}\setminus m\mathbb{Z}_{mgw}$ and the size of ${\cal B}'_j$ is $uw$ for any $0\leq j < M$.

Since ${\cal B}_j$ is an $(mg,g,K_j,\lambda)$-CRDP of size $u$, the difference list $\Delta({\cal B}_j)$ contains each element of $\mathbb{Z}_{mg}\setminus m\mathbb{Z}_{mg}$ at most $\lambda$ times and no element of $m\mathbb{Z}_{mg}$ occurs. Simple computation shows that
\[
\begin{array}{l}
\Delta({\cal B}'_j)=\bigcup\limits_{0\leq i<u, \atop 0\leq s<w}\Delta(B_{(i,s)}^j) \\
=\bigcup\limits_{0\leq i<u}\{a-b+cmg:~ a\neq b \in B_i^j, \ 0\leq c < w\}\\
=\bigcup\limits_{\tau\in \Delta({\cal B}_j)}(mg\mathbb{Z}_{mwg}+\tau),
\end{array}
\]
consequently, the difference list $\Delta({\cal B}'_j)$ contains each element of $\mathbb{Z}_{mgw}\setminus m\mathbb{Z}_{mgw}$ at most $\lambda$ times and no element of $m\mathbb{Z}_{mgw}$ occurs. So, each ${\cal B}'_j$ is an $(mgw,gw,K_j,\lambda)$-CRDP of size $uw$.

Since $\Delta_E({\cal B}_j, {\cal B}_{j'})$ contains each residue of $\mathbb{Z}_{mg}\setminus m\mathbb{Z}_{mg}$ at most $\lambda$ times and no element of $m\mathbb{Z}_{mg}$ occurs for $0\leq j\neq j' < M$, we get
\[
\begin{array}{l}
\Delta_E({\cal B}'_j, {\cal B}'_{j'})=\bigcup\limits_{0\leq i<u}\bigcup\limits_{0\leq s<w}\Delta_E(B_{(i,s)}^j, B_{(i,s)}^{j'}) \\
=\bigcup\limits_{0\leq i<u}\{b-a+cmg:~ (a,b) \in B_i^j\times B_i^{j'}, \ 0\leq c < w\}\\
=\bigcup\limits_{\tau\in \Delta_E({\cal B}_j, {\cal B}_{j'})}(mg\mathbb{Z}_{mwg}+\tau),
\end{array}
\]
consequently, the difference list $\Delta_E({\cal B}'_j, {\cal B}'_{j'})$ contains each residue of $\mathbb{Z}_{mg}\setminus m\mathbb{Z}_{mg}$ at most $\lambda$ times and no element of $m\mathbb{Z}_{mg}$. This completes the proof. 
\end{IEEEproof}


Combining Theorem \ref{HFHS} with Theorem \ref{optimal 3p} establishes the following corollary.

\begin{corollary}\label{optimal 3p_1p_2}
There exists an optimal $(n,2,4; \frac{n+1}{4})$-FHS set for all $n=3p_1p_2\cdots p_u$ with $n\not \equiv 0 \pmod {25}$ and each $p_j$ $\equiv 1 \pmod {4}$ being a prime.
\end{corollary}

\begin{IEEEproof} We first prove that there exists a partition-type $(n,\{\{3,4\},\{3,4\}\},4)$-BNCDP of size $\frac{n+1}{4}$ over $\mathbb{Z}_{n}$ by induction on $u$. Without loss of generality, let $p_1\leq p_2\leq \cdots \leq p_u$.

For $u=1$, the assertion holds by Theorem \ref{optimal 3p}. Assume that the assertion holds for $u=r$ and consider $u=r+1$. Deleting the block $\{(0,0),(1,0),(2,0)\}$ from ${\cal B}_t$ in the proof of Theorem \ref{optimal 3p} where $0\leq t<2$, we obtain a $(3p_{1},3,\{\{4\},\{4\}\},4)$-BNCRDP of size $\frac{3p_{1}-3}{4}$, $\{{\cal B}'_0,{\cal B}'_1\}$ such that each ${\cal B}'_t$ is a partition of $\mathbb{Z}_{3p_{1}}\setminus p_{1}\mathbb{Z}_{3p_{1}}$. Since $3p_1p_2\cdots p_{r+1}\not\equiv 0\pmod {25}$ and each $p_j\equiv 1\pmod 4$ is a prime, we have $p_j\geq 13$ for $2\leq j\leq r+1$.
By Lemma \ref{CDM} there exists a homogeneous $(p_2\cdots p_{r+1},8,1)$-CDM. Since $8=|B_i^0|+|B_i^1|$ where $B_i^0 \in {\cal B}'_0$ and $B_i^1 \in {\cal B}'_1$ for $0\leq i<\frac{3p_1-3}{4}$, applying Theorem \ref{HFHS} yields a $(3p_1\cdots p_{r+1},3 p_2\cdots p_{r+1},\{\{4\},\{4\}\},4)$-BNCRDP  of size $\frac{3p_1\cdots p_{r+1}- 3p_2\cdots p_{r+1}}{4}$ over $\mathbb{Z}_{3p_1\cdots p_{r+1}}$ such that each CRDP is a partition of $\mathbb{Z}_{3p_1\cdots p_{r+1}}\setminus p_{1}\mathbb{Z}_{3p_1\cdots p_{r+1}}$.
By induction hypothesis there exits a partition-type $(3p_2\cdots p_{r+1},\{\{3,4\},\{3,4\}\},4)$-BNCDP over $\mathbb{Z}_{3p_2\cdots p_{r+1}}$ of size $\frac{3p_2\cdots p_{r+1}}{4}$, applying Lemma \ref{BNCRDP=>BNCDP} then yields a partition-type $(3p_1\cdots p_{r+1},\{\{3,4\},\{3,4\}\},4)$-BNCDP of size $\frac{3p_1\cdots p_{r+1}+1}{4}$ over $\mathbb{Z}_{3p_1\cdots p_{r+1}}$. So, the conclusion holds by induction.

By Theorem \ref{FHS sets}, there exists an $(n,2,4; \frac{n+1}{4})$-FHS set. It remains to show that such an FHS set is optimal.

Since $k=\left\lfloor\frac{n}{l}\right\rfloor=\left\lfloor\frac{n}{\frac{n+1}{4}}\right\rfloor=3$, we get $\epsilon = n-3\times \frac{n+1}{4}=\frac{n-3}{4}$ and $\epsilon M-l=2\times \frac{n-3}{4}-\frac{n+1}{4}=\frac{n-7}{4}>0.$ By Corollary \ref{bounds}, this FHS set is optimal.

\end{IEEEproof}

It is worth pointing out that the construction of FHS sets in \cite[Theorem 3.10]{GMY2009} is just a special case of Theorem \ref{HFHS}. The construction of FHS sets in \cite[Theorem 3.10]{GMY2009} require that one frequency of an $(n,M,\lambda;l)$-FHS set appears in a fixed position and other frequencies appear in different positions. Such a FHS set is in fact equivalent to an $(n,1,\{K_0,\ldots,K_{M-1}\},\lambda)$-BNCRDP of size $l-1$.
Applying Theorem \ref{HFHS} with a homogeneous $(w,t,1)$-CDM yields a $(nw,w,\{K_0,\ldots,K_{M-1}\},\lambda)$-BNCRDP of size $(l-1)w$. Further, applying Lemma \ref{BNCRDP=>BNCDP} with a partition-type $(w,\{K_0',\ldots,K_{M-1}'\},\lambda)$-BNCDP of size $r$ yields a partition-type $(nw,\{K_0\cup K_0',\ldots,K_{M-1}\cup K_{M-1}'\},\lambda)$-BNCDP of size $(l-1)w+r$, which corresponds to an $(nw,M,\lambda;(l-1)w+r)$-FHS set.

\begin{theorem}\cite{GMY2009}\label{difference matrix}
Assume that $S$ is an $(n,M,\lambda;l)$-FHS set in which one frequency appears in a fixed position, say the $0$th position, and each of the other frequencies appears in different non-$0$th positions of the $M$ FHSs of $S$. Assume also that $T$ is a $(w,M,\lambda;r)$-FHS set. If there exists a homogeneous $(w,t,1)$-CDM over $\mathbb{Z}_w$, where $t$ is the maximum number of total occurrences that frequencies appear in all the $M$ FHSs of $S$, then there also exists an $(nw,M,\lambda;(l-1)w+r)$-FHS set.
\end{theorem}

When we replace the BNCRDP in Theorem \ref{HFHS} with a partition-type BNCDP, the same procedure yields a new partition-type BNCDP, which is stated in terms of FHS set below.
Since the proof is similar to that of Theorem \ref{HFHS}, we omit it here.
\begin{theorem}\label{difference matrix 2}
Assume that $S$ is an $(n,M,\lambda;l)$-FHS set. If there exists a homogeneous difference matrix $(w,t,1)$-DM over $\mathbb{Z}_w$, where $t$ is the maximum number of total occurrences that frequencies appear in all the $M$ FHSs of $S$, then there also exists an $(nw,M,\lambda;lw)$-FHS set.
\end{theorem}

Applying Theorem \ref{difference matrix 2} and Lemma \ref{CDM} gives the following corollary.
\begin{corollary}\label{nv}
Assume that $S$ is an $(n, M, \lambda; l)$-FHS set. Let $w$ be an odd integer and let $q_1$ be the least prime divisor of $w$.
If $t<q_1$, where $t$ is the maximum number of total occurrences that frequencies appear in all the $M$ FHSs of $S$, then there exists an $(nw,M, \lambda; wl)$-FHS set.
\end{corollary}

Remark: Compared with the Construction A in \cite{CGY2014}, the construction for FHS sets from Corollary \ref{nv} does not require the constraint $gcd(w,n)=1$.
As noted in \cite{CGY2014}, the resultant $(nw,M,\lambda;lw)$-FHS set is optimal if the $(n, M, \lambda; l)$-FHS set is optimal.

\begin{lemma}\cite{ZTPP2011}\label{q^n}
Let $m,u$ be positive integers with $u< m$, $q$ a prime power and let $d$ be a positive integer such that $d|q-1$ and gcd$(d,m)=1$. Then there exists an optimal $(\frac{q^m-1}{d},d, \frac{q^{m-u}-1}{d}; q^u)$-FHS set.
\end{lemma}

From the construction in \cite{ZTPP2011}, the maximum number of total occurrences that frequencies appear in all FHSs of the $(\frac{q^m-1}{d},d, \frac{q^{m-u}-1}{d}; q^u)$-FHS set is $q^{m-u}$. When the least prime factor of $w$ is greater than $q^{m-u}$, applying Corollary \ref{nv} yields a $(\frac{w(q^m-1)}{d},d, \frac{q^{m-u}-1}{d}; wq^u)$-FHS set, which is optimal based on the remark of Corollary \ref{nv}.
Hence, we have the following  corollary.

\begin{corollary}\label{v(q^n-1)}
Let $m,u$ be positive integers with $u< m$, $q$ a prime power and let $d$ be a positive integer such that $d|q-1$ and gcd$(d,m)=1$. Let $w$ be an odd integer whose the least prime factor is greater than $q^{m-u}$. Then there exists an optimal $(\frac{w(q^m-1)}{d},d, \frac{q^{m-u}-1}{d}; wq^u)$-FHS set.
\end{corollary}

Corollary \ref{v(q^n-1)} can yield some optimal FHS sets that can not be obtained by Construction A in \cite{CGY2014}. For example, take $w=13, q=3,$ $m=3, u=1$ and $d=2$, we obtain an optimal $(169,2,4;39)$-FHS set.

In the sequel, we shall make use of Theorem 4.7 to obtain new optimal FHS sets.

Let $v>1$ be an odd integer. An element $g\in U(\mathbb{Z}_v)$ is called a primitive root modulo $v$ if its multiplicative order modulo $v$ is $\varphi(v)$, where $\varphi(v)$ denotes the Euler function which counts the number of positive integers less than and coprime to $v$. It is well known that for an odd prime $p$, there exists an element $g$ such that $g$ is a primitive root modulo $p^t$ for all $t\geq 1$ \cite{A1976}.

Let $v$ be an odd integer of the form $v=p_1^{m_1}p_2^{m_2}\cdots p_s^{m_s}$ for $s$ positive integers $m_1,m_2,\ldots,m_s$ and $s$ distinct primes $p_1,p_2,\ldots,p_s$. Let $e>1$ be a common factor of $p_1-1,p_2-1,\ldots,p_s-1$. Define $f=\min \{\frac{p_i-1}{e}:~ 1\leq i\leq s\}$. For each $i$ with $1\leq i \leq s$, let $g_i$ be a primitive root modulo $p_i^t$ for all $t\geq1$. By the Chinese Remainder Theorem, there exist unique elements $g, a\in U(\mathbb{Z}_v)$ such that
\[
\begin{array}{l}
\vspace{0.1cm} g\equiv g_i^{f_ip_i^{m_i-1}}\pmod {p_i^{m_i}}\ {\rm for}\ 1\leq i\leq s, \\
a\equiv g_i \pmod {p_i^{m_i}}\ {\rm for}\ 1\leq i\leq s,
\end{array}
\]
then the multiplicative order of $g$ modulo $v$ is $e$, the list of differences arising from $G=\{1,g,\ldots, g^{e-1}\}$ is a subset of $U(\mathbb{Z}_v)$ and $a^tg^c-g^{c'}\in U(\mathbb{Z}_v)$ for $1\leq t<f$ and $0\leq c,c'<e$.

For $x,y\in \mathbb{Z}_v\setminus\{0\}$, the binary relation $\sim$ defined by $x\sim y$ if and only if there exists a $g' \in G$ such that $xg'=y$ is an equivalence relation over $\mathbb{Z}_v\setminus\{0\}$. Then its equivalence classes are the subsets $xG, x\in \mathbb{Z}_v\setminus\{0\},$ of $\mathbb{Z}_v$. Denote by $R$ a system of distinct representatives for the equivalence classes modulo $G$ of $\mathbb{Z}_v\setminus\{0\}$. For $0\leq t < f, r\in R$, set
$$B_{r}^t=\{ra^tg^j:~0\leq j < e\}$$ and
define $${\cal B}_t=\{B_{r}^t:~ r\in R\}.$$

Since $\bigcup\limits_{r\in R}B_r^t=\{ra^tg^j:~r\in R,\ 0\leq j< e\}=\mathbb{Z}_v\setminus\{0\}$, we get
\[
\begin{array}{l}
\vspace{0.20cm}\Delta({\cal B}_t)=\bigcup\limits_{r\in R}\Delta(\{ra^tg^j:~ 0\leq j< e \})\\
\vspace{0.20cm}\hspace{1cm}=\bigcup\limits_{r\in R}\{ra^t(g^{j'}-g^j):~ 0\leq j' \neq j < e\}\\
\vspace{0.20cm}\hspace{1cm}=\bigcup\limits_{r\in R}\{ra^tg^j(g^c-1):~0\leq j < e, 1\leq c < e \}\\
\vspace{0.20cm}\hspace{1cm}=(e-1)[\mathbb{Z}_v\setminus \{0\}].
\end{array}
\]
Thus, ${\cal B}_t$ is a $(v,1,e,e-1)$-CRDP for $0\leq t< f$.

For $0\leq t\neq t'< f$, since $a^{t'-t}g^c-1\in U(\mathbb{Z}_v)$ and $\{ra^{t}g^j:~r\in R,\ 0\leq j< e\}=\mathbb{Z}_v\setminus\{0\}$, we get
\[
\begin{array}{l}
\vspace{0.2cm}\Delta_E({\cal B}_t,{\cal B}_{t'})=\bigcup\limits_{r\in R}\Delta_E(B_{r}^t,B_{r}^{t'})\\
\vspace{0.2cm}\hspace{1.6cm}=\bigcup\limits_{r\in R}\{ra^{t'}g^{j'}-ra^tg^{j}:~ 0\leq j, j' < e \}\\
\vspace{0.2cm}\hspace{1.6cm}=\bigcup\limits_{c=0}^{e-1}\bigcup\limits_{r\in R}\{ra^{t}g^j(a^{{t'}-t}g^c-1):~0\leq j < e \}\\
\hspace{1.6cm}=e[\mathbb{Z}_v\setminus \{0\}].
\end{array}
\]
It follows that $\{{\cal B}_t:~0\leq t<f\}$ is a $(v,1,\{K_0,K_1,\ldots,K_{f-1}\},e)$-BNCRDP with each ${\cal B}_t$ being a partition of $\mathbb{Z}_{v}\setminus\{0\}$, where $K_0=K_1=\cdots=K_{f-1}=\{e\}$.

The discussion above establishes the following lemma.

\begin{lemma}\label{BNCRDP 2}
Let $v$ be an odd integer of the form $p_1^{m_1}p_2^{m_2}\cdots p_s^{m_s}$ for $s$ positive integers $m_1,m_2,\ldots,m_s$ and $s$ distinct primes $p_1,p_2,\ldots,p_s$. Let $e>1$ be a common factor of $p_1-1, p_2-1, \cdots, p_s-1$ and let $f=\min\{\frac{p_i-1}{e}:~1\leq i\leq s\}$. Then there exists a $(v, 1, \{K_0,K_1,\ldots,K_{f-1}\}, e)$-BNCRDP of size $\frac{v-1}{e}$ such that each CRDP is a partition of $\mathbb{Z}_{v}\setminus\{0\}$, where $K_0=K_1=\cdots=K_{f-1}=\{e\}$.
\end{lemma}

Adding the block $\{0\}$ to ${\cal B}_t$ for each $0\leq t<f$, the new collection is a partition-type $(v, \{K_0',\ldots,K_{f-1}'\}, e)$-BNCDP of size $\frac{v-1}{e}+1$  where $K_0'=\cdots=K_{f-1}'=\{1,e\}$, which corresponds to a $(v,f,e;\frac{v-1}{e}+1)$-FHS set. Such an FHS set can also be obtain from Construction A in \cite{ZCTY2013} by using generalized cyclotomy. In comparison, our method is quite neat and more clear to understand.

\begin{corollary}\cite{ZCTY2013}\label{BNCDP 3}
Under the hypotheses of Lemma \ref{BNCRDP 2}, 
there exists a partition-type $(v, \{K_0',\ldots,K_{f-1}'\}, e)$-BNCDP
of size $\frac{v-1}{e}+1$,  where $K_0'=\cdots=K_{f-1}'=\{1,e\}$.
\end{corollary}

\begin{theorem}\label{optimal v_1v_2}
Let $w$ be an odd integer of the form $q_1^{n_1}q_2^{n_2}\cdots q_t^{n_t}$
for $t$ positive integers $n_1,n_2,\ldots,n_t$ and $t$ primes $q_1,q_2,\ldots,q_t$ such that $q_1<\cdots<q_t$,
and let $e'$ be a positive integer such that $e'|q_i-1$ for $1\leq i \leq t$.
Let parameters $v$, $p_1$, $e$ and $f$ be the same as those in the hypotheses of Lemma \ref{BNCRDP 2}.  If $2\leq e'\leq e$, $q_1\geq p_1>2e$ and $v \geq e^2$, then there exists an optimal
$(vw,\frac{p_1-1}{e},e;\frac{v-1}{e}w+\frac{w-1}{e^{'}}+1)$-FHS set.
\end{theorem}

\begin{IEEEproof}
By Lemma \ref{BNCRDP 2}, there exists a $(v,1,\{K_0,\ldots, K_{\frac{p_1-1-e}{e}}\},e)$-BNCRDP of size $\frac{v-1}{e}$, $\{{\cal B}_j:~ 0\leq j<\frac{p_1-1}{e}\}$ such that each ${\cal B}_j$ is a partition of $\mathbb{Z}_{v}\setminus\{0\}$, where
$K_0=\cdots=K_{\frac{p_1-1-e}{e}}=\{e\}$. Since $p_1\leq q_1$,  by Lemma \ref{CDM} there exists a homogeneous $(w, p_1-1, 1)$-CDM over
$\mathbb{Z}_{w}$. Since $p_1-1= \sum\limits_{j=0}^{\frac{p_1-1}{e}-1}|B_i^j|$ for $0\leq i<\frac{v-1}{e}$, applying Theorem \ref{HFHS} yields a $(vw,w,\{K_0,\ldots, K_{\frac{p_1-1-e}{e}}\},e)$-BNCRDP of size
$\frac{vw-w}{e}$ such that each CRDP is a partition of
$\mathbb{Z}_{vw}\setminus v\mathbb{Z}_{vw}$. By Corollary \ref{BNCDP 3}, there exists a partition-type
$(w, \{K'_0,\ldots, K'_{\frac{q_1-1-e'}{e'}}\}, e')$-BNCDP of size $\frac{w-1}{e^{'}}+1$, where
$K'_0=\cdots=K'_{\frac{q_1-1-e'}{e'}}=\{1,e'\}$. Since $\frac{q_1-1-e'}{e'}\geq \frac{p_1-1-e}{e}$, applying Lemma \ref{BNCRDP=>BNCDP}
we obtain a partition-type $(vw,\{K^1_0,\ldots, K^1_{\frac{p_1-1-e}{e}}\},e)$-BNCDP of size $\frac{v-1}{e}w+\frac{w-1}{e^{'}}+1$,
where $K^1_0=\cdots=K^1_{\frac{p_1-1-e}{e}}=\{1,e,e'\}$. Therefore, by Theorem \ref{FHS sets} there is a
$(vw,\frac{p_1-1}{e},e;\frac{v-1}{e}w+\frac{w-1}{e^{'}}+1)$-FHS set.
It remains to prove that such an FHS set is optimal.

Since $v \geq e^2$, simple computation shows that $e-1\leq  \frac{vw}{\frac{v-1}{e}w+\frac{w-1}{e^{'}}+1} <e$. Consequently,
\[
\begin{array}{l}
\vspace{0.2cm}k=\left\lfloor \frac{vw}{\frac{v-1}{e}w+\frac{w-1}{e^{'}}+1} \right\rfloor =e-1, \ {\rm and}\\
 \vspace{0.2cm} \epsilon=n-kl=vw-(e-1)(\frac{v-1}{e}w+\frac{w-1}{e^{'}}+1)
\\  \vspace{0.2cm} \hspace{0.2cm} =\frac{vw-w}{e}-(e-1)\frac{w-1}{e^{'}}+ w-e+1.
\end{array}
\]
Since $w\geq p_1 \geq 2e$, $v \geq e^2$ and $e\geq e'\geq 2$, we get

\[
\begin{array}{l}
 \vspace{0.2cm} \epsilon M-l\geq 2\epsilon-l
\\  \vspace{0.2cm}  \hspace{1.1cm} =(\frac{vw-w}{e}+\frac{w-1}{e^{'}})-2e \frac{w-1}{e^{'}}+(2w-2e+1)
\\  \vspace{0.2cm} \hspace{1.1cm}\geq \frac{vw-1}{e}-e(w-1)+(2w-2e+1)
\\  \vspace{0.2cm} \hspace{1.1cm} > \frac{(v-e^2)w+e^2-1}{e} >0.
\end{array}
\]
By Corollary \ref{bounds}, this FHS set is optimal. This completes the proof.
\end{IEEEproof}

Remark: Compared with the construction A in \cite{ZCTY2013}, the construction for FHS set from Theorem \ref{optimal v_1v_2} does not
require the constraint $e|gcd(p_1-1,p_2-1,\ldots,p_s-1, q_1-1,\ldots,q_t-1)$. From Theorem \ref{optimal v_1v_2}, we can obtain many FHS
sets with new and flexible parameters.

Similar to the construction of FHS sets in Theorem \ref{optimal v_1v_2}, we can obtain more  FHS sets by using known FHS sets,
Theorem \ref{HFHS} and Theorem \ref{BNCRDP=>BNCDP}. Here, we give another example.

\begin{lemma} \cite{CHY2009} \label{kn}
Assume that there exists an $(n,M,\lambda;l)$-FHS set. Then, for any integer $t$ with $1\leq t \leq M$, there exists a
$(tn,\left\lfloor\frac{M}{t}\right\rfloor,t\lambda;l)$-FHS set.
\end{lemma}


\begin{theorem}\label{optimal qv}
Let $p,p'$ be primes and let $m,a,b$ be positive integers such that $p^m-1=ab$ and $p'(b+1)\leq a$.
Let parameters $v,p_1,e,f$ and $w,q_1,e'$ be the same as those in the hypotheses of Theorem \ref{optimal v_1v_2}.
If $p^m-1<p_1\leq q_1$ and $e\leq b$,
then there exists an optimal $(p'vw(p^m-1),\lfloor \frac{a}{p'}\rfloor,p'b;avw+\frac{(v-1)w}{e}+\frac{w-1}{e^{'}}+1)$-FHS set.
\end{theorem}

\begin{IEEEproof}
Let $M=\left\lfloor \frac{a}{p'}\right\rfloor$. Since gcd$(p',p^m-1)=1$, we have that $\mathbb{Z}_{p'(p^m-1)}$ is isomorphic to $\mathbb{Z}_{p'}\times \mathbb{Z}_{p^m-1}$.
From the construction of $(p'(p^m-1), \left\lfloor \frac{a}{p'}\right\rfloor ,p'b; a+1)$-FHS set in \cite{RFZ2014}, there exists a base block
$\{(i,c):~0\leq i<p'\}$ for some $c$ in the corresponding partition-type BNCDP over $\mathbb{Z}_{p'}\times \mathbb{Z}_{p^m-1}$. We translate
blocks to obtain a $(p'(p^m-1), p', \{K_0,\ldots, K_{M-1}\}, p'b)$-BNCRDP of size $a$ with each CRDP being a partition of
$\mathbb{Z}_{p'(p^m-1)}\setminus (p^m-1)\mathbb{Z}_{p'(p^m-1)}$, where $K_0=\cdots= K_{M-1}=\{p'b,p'(b-1)\}$.

Since $p^m-1<p_1\leq q_1$, there exists a homogeneous $(vw, p^m-1, 1)$-CDM over $\mathbb{Z}_{vw}$ by Lemma \ref{CDM}. Since the block size is at most $p'b$, the sum of the cardinalities of $M$ blocks is at most $Mp'b$, which is not greater than $p^m-1$. Applying Theorem \ref{HFHS} yields a $(p'vw(p^m-1),p'vw, \{K_0,\ldots, K_{M-1}\}, p'b)$-BNCRDP of size $avw$ such that each CRDP is a partition of $\mathbb{Z}_{p'vw(p^m-1)}\setminus (p^m-1)\mathbb{Z}_{p'vw(p^m-1)}$.  Since there exists a $(vw,\frac{p_1-1}{e},e;\frac{vw-w}{e}+\frac{w-1}{e'}+1)$-FHS set by Theorem \ref{optimal v_1v_2}, applying Lemma \ref{kn} yields a $(p'vw,\lfloor\frac{p_1-1}{p'e}\rfloor,p'e;\frac{vw-w}{e}+\frac{w-1}{e'}+1)$-FHS set.
By Theorem \ref{FHS sets}, there is a partition-type $(p'vw,\{K'_0,\ldots, K'_{M'-1}\},p'e)$-BNCDP of size $\frac{vw-w}{e}+\frac{w-1}{e'}+1$, where $M'=\left\lfloor \frac{p_1-1}{p'e}\right\rfloor$. Since $2\leq e \leq b$ and $p_1>p^m-1$, it holds that $M'=\left\lfloor\frac{p_1-1}{p'e}\right\rfloor\geq \left\lfloor\frac{ab}{p'e}\right\rfloor\geq \left\lfloor\frac{a}{p'}\right\rfloor=M$. Applying Lemma \ref{BNCRDP=>BNCDP} and Theorem \ref{FHS sets}, we obtain a $(p'vw(p^m-1),\lfloor \frac{a}{p'}\rfloor,p'b;avw+\frac{vw-w}{e}+\frac{w-1}{e'}+1)$-FHS set.
It remains to prove that this FHS set is optimal.

Since $2\leq e' \leq e$, $p'(b+1)\leq a$ and $p^m-1=ab$, we have
\[
\begin{array}{l}
\vspace{0.2cm}(p'b-1)(avw+\frac{vw-w}{e}+\frac{w-1}{e'}+1)\\
\vspace{0.2cm} \leq (p'b-1)(avw+\frac{vw-1}{e}+\frac{vw-1}{e'}+1) \\
\vspace{0.2cm} \leq (p'b-1)(avw+vw) \\
\vspace{0.2cm} = p'(p^m-1)vw+(p'b-1-a)vw \\
\vspace{0.2cm} \leq p'(p^m-1)vw,\ \ {\rm and } \\
p'vw(p^m-1)< p'b(avw+\frac{vw-w}{e}+\frac{w-1}{e'}+1).
\end{array}
\]
Hence,
\[
\begin{array}{l}
k=\left\lfloor \frac{p'vw(p^m-1)}{avw+\frac{vw-w}{e}+\frac{w-1}{e'}+1} \right\rfloor=p'b-1, \ {\rm and} \\

\end{array}
\]
\[
\begin{array}{l}
\vspace{0.2cm} \epsilon=p'vw(p^m-1)\\
\vspace{0.2cm}\hspace{0.8cm} -(p'b-1)(avw+\frac{vw-w}{e}+\frac{w-1}{e'}+1)\\
\vspace{0.2cm} \hspace{0.3cm}= avw+(1-p'b) (\frac{vw-w}{e}+\frac{w-1}{e'}+1).
\end{array}
\]
Since $b$ is a positive integer, $M=\left\lfloor \frac{a}{p'}\right\rfloor$ and $p'(b+1)\leq a$, we have
$M\geq b+1$ and  $(M-1)a \geq M(p'b-1)+1$. Then,
\[
\begin{array}{l}
\vspace{0.2cm} \epsilon M-l= M(avw+(1-p'b) (\frac{vw-w}{e}+\frac{w-1}{e'}+1))\\
\vspace{0.2cm} \hspace{1.5cm} -(avw+\frac{vw-w}{e}+\frac{w-1}{e'}+1)\\
\vspace{0.2cm} =(M-1)avw-(M(p'b-1)+1)(\frac{vw-w}{e}+\frac{w-1}{e'}+1) \\
\vspace{0.2cm} \geq(M-1)avw-(M-1)a(\frac{vw-w}{e}+\frac{w-1}{e'}+1) \\
\vspace{0.2cm} =(M-1)a(vw-(\frac{vw-w}{e}+\frac{w-1}{e'}+1))\\
\vspace{0.2cm} \geq (M-1)a(vw-\frac{vw+1}{2}) \\
 >0.
\end{array}
\]
By Corollary \ref{bounds}, this FHS set is optimal.
\end{IEEEproof}

If $p'(b+1)>a$, then the $(p'(p^m-1), \left\lfloor \frac{a}{p'}\right\rfloor ,p'b; a+1)$-FHS set constructed in \cite{RFZ2014}  may not be optimal. However,  the resultant FHS set constructed in the proof of Theorem \ref{optimal qv} may be optimal with respect to the Peng-Fan bound (\ref{Bound 6}). For example, take $p=5,p'=3, m=2,a=8$ and $b=3$, we obtain a $(72,2,9;9)$-FHS set. It is easy to see that this FHS set is not optimal with respect to the Peng-Fan bounds. Let $p,q$ be two primes such that $p\equiv q\equiv 1\pmod 3$ and $q>p \geq 73$. When  we take $e=e'=3$ in the proof of  Theorem \ref{optimal qv}, we obtain a $(72pq, 2, 9;\frac{25pq+2}{3})$-FHS set. Clearly, this new FHS set is optimal with respect to the Peng-Fan bound (\ref{Bound 6}).

\section{Concluding Remarks}  %
\label{concl}                               %

We showed an algebraic construction, two direct constructions and recursive constructions for FHS sets. From these constructions, we obtained many infinitely  families of new optimal FHS sets with respect to the Peng-Fan bound (4). Our combinatorial constructions generalized the previous methods, the recursive construction for BNCDPs in \cite{GMY2009} became a special case of Theorem \ref{HFHS}, the constraint $gcd(w,n)=1$ of Construction A in \cite{CGY2014} was removed, the existence proof of a $(v,f,e;\frac{v-1}{e}+1)$-FHS set in \cite{ZCTY2013} was simplified by using cyclotomic cosets. Compared with the previous extension methods in \cite{CGY2014}, \cite{CHY2009}, \cite{GMY2009} and \cite{ZCTY2012}, our recursive constructions gave new optimal FHS sets for much more general cases. 

\ifCLASSOPTIONcaptionsoff
  \newpage
\fi

\end{document}